\documentclass[twocolumn,showpacs,superscriptaddress,preprintnumbers,amsmath,amssymb,fleqn]{revtex4}
\usepackage{graphicx}
\usepackage{dcolumn}
\usepackage{bm}

\begin{document}

\preprint{INHA-NTG-08/09}

\title{Non-Fermi liquid from confinement in doped Mott insulators}

\author{Ki-Seok Kim}

\affiliation{Asia Pacific Center for Theoretical Physics, Hogil
Kim Memorial building 5th floor, POSTECH, Hyoja-dong, Namgu,
Pohang 790-784, Korea}

\author{Hyun-Chul Kim}

\affiliation{Department of Physics, Inha University, Incheon
402-751 Korea}

\date{\today}

\begin{abstract}
A phenomenological description for confinement of fractionalized
excitations is proposed in the gauge theory approach for doped
Mott insulators. Introducing the Polyakov-loop parameter into an
SU(2) gauge theory for the t-J model, we show that electron
excitations emerge below the so-called coherence temperature,
resulting from confinement of spinons and holons via the formation
of the Polyakov loop. Remarkably, such confined electrons turn out
to exhibit non-Fermi liquid physics without quantum criticality,
yielding the electric resistivity in quantitative agreement with
experimental data. The Higgs phase is not allowed due to
confinement, suggesting a possible novel mechanism of
superconductivity in the strong coupling approach.
\end{abstract}

\pacs{71.10.Hf, 71.10.-w, 71.10.Fd, 71.30.+h}

\maketitle

Research on strongly correlated electrons gives rise to crisis in
two cornerstones of modern theory of metals, Landau Fermi liquid
theory and Landau-Ginzburg-Wilson framework for phase
transitions~\cite{QCP_Review}. Gauge theory formulation has been
proposed to incorporate strong correlations, from which spin
liquid physics, being described by fractionalized
excitations~\cite{Spin_Liquid}, emerges as a main prediction.
Although this strong coupling approach explains the anomalous charge
and spin dynamics of high T$_{c}$ cuprates at high temperatures,
it fails to grasp why coherent electron excitations are
observed at low temperatures~\cite{Lee_Nagaosa_Wen}. This problem
was often claimed to relate to confinement in the gauge theory,
being regarded as an elusive one at the present technology.

Confinement is the most salient and difficult feature of quantum
chromodynamics (QCD)~\cite{Confinement_Review}. Though lattice QCD
simulations shed some light on it~\cite{Lattice_Review}, it still
defies any complete understanding. Recently, Fukushima proposed the
Polyakov-loop extended Nambu-Jona-Lasinio (PNJL) model, where the
Polyakov-loop parameter, measuring an effective potential for creation
of static single quark~\cite{Ployakov_Loop}, is introduced into the
NJL model for the characterization of
confinement~\cite{Fukushima}. This effective model has merits, in
particular, in studying the quark matter at high temperatures, since
both spontaneously broken chiral symmetry (SB$\chi$S) and confinement
are described on an equal footing, being consistent with the lattice
simulation~\cite{PNJL_Models}.

In this Letter we introduce the PNJL scheme to the gauge theory
approach for doped Mott insulators for the first time. Considering
the Polyakov-loop parameter in an SU(2) slave-boson
theory~\cite{Lee_Nagaosa_Wen} of the t-J model, we show that
electron excitations emerge below the so-called coherence
temperature, being ascribed to confinement of spinons and holons via
the formation of the Polyakov loop. Remarkably, such confined
electrons turn out to exhibit non-Fermi liquid physics without
quantum criticality, fitting experimental data for the electric
resistivity quantitatively well. The Higgs phase given by the holon
condensation is not allowed due to confinement, implying a novel
mechanism of superconductivity, which is distinguished from the
previous gauge theory approaches based on deconfinement.

We start from the SU(2) slave-boson representation of the t-J
model \cite{Lee_Nagaosa_Wen}
\begin{eqnarray}
Z &=& \int D \psi_{i\alpha} D
h_{i} D U_{ij} D a_{i0}^{k} e^{- \int_{0}^{\beta} d \tau L} , \cr
L &=& \frac{1}{2} \sum_{i} \psi_{i\alpha}^{\dagger}
(\partial_{\tau} - i a_{i0}^{k}\tau_{k})\psi_{i\alpha} + J
\sum_{\langle i j \rangle} (
\psi_{i\alpha}^{\dagger}U_{ij}\psi_{j\alpha} \cr
&+& H.c.) +
\sum_{i} h_{i}^{\dagger}(\partial_{\tau} - \mu  - i
a_{i0}^{k}\tau_{k}) h_{i} \cr
&& + t \sum_{\langle i j \rangle} (
h_{i}^{\dagger}U_{ij}h_{j} + H.c.) + J \sum_{\langle i j
\rangle} \mathbf{tr}[U_{ij}^{\dagger}U_{ij}] ,
\label{eq:1}
\end{eqnarray}
where both spinon $\psi_{i\sigma}^{\dagger} = \left(
\begin{array}{cc} f_{i\sigma}^{\dagger} &
\epsilon_{\sigma\sigma'} f_{i\sigma'} \end{array} \right)$ and
holon $h_{i}^{\dagger} = \left(
\begin{array}{cc} b_{i1}^{\dagger} & b_{i2}^{\dagger} \end{array}
\right)$ fields are given by doublets, carrying spin and charge
quantum numbers of an electron, respectively. The order parameter
matrix $U_{ij} = \left( \begin{array}{cc} - \chi_{ij}^{\dagger} &
\Delta_{ij} \\ \Delta_{ij}^{\dagger} & \chi_{ij} \end{array}\right)$
comes from the standard decomposition for interactions, where
$\chi_{ij}$ and $\Delta_{ij}$ are associated with particle-hole and 
particle-particle channels, respectively. Equation~(\ref{eq:1}) should
be regarded as one reformulation of the t-J model, decomposing an
electron field into spinon and holon fields given by $c_{i\sigma}
= \frac{1}{\sqrt{2}} h_{i}^{\dagger} \psi_{i\sigma}$, where the
Gutzwiller projection is replaced with the exact integration of
$a_{i0}^{k}$.

Employing the mean-field approximation for $U_{ij}$ and
$a_{i0}^{k}$, Wen and Lee~\cite{Lee_Nagaosa_Wen} found the phase
diagram of the effective theory represented by Eq.~(\ref{eq:1}) in the
$(\delta, T)$ plane with a fixed $J/t$, where $\delta$ denotes a hole
concentration and $T$ stands for temperature. The
optimally hole-doped region at high temperatures is described by
$U_{ij}^{SM} = - i \chi I$ and $\langle h_{i} \rangle = 0$ with
$a_{i0}^{k} = 0$, called the strange metal (SM) phase, where
spinons form a large Fermi surface, but only incoherent electron
spectra are observed. The under-doped region at intermediate
temperatures is characterized by $U_{ij}^{SF} = - \sqrt{\chi^{2} +
\Delta^{2}} \tau_{3} \exp[i(-1)^{i_{x} + i_{y}} \Phi \tau_{3}]$
with $\Phi = \tan^{-1}\Bigl( \frac{\Delta}{\chi} \Bigr)$, $\langle
h_{i} \rangle = 0$, and $a_{i0}^{k} = 0$, called the staggered
flux (SF) phase, where spinons have Dirac spectrum due to the
staggered internal flux $\Phi$, but coherent electrons are not
seen as the SM phase. Because the electron spectrum exhibits its
spectral gap except for Dirac points, this SF state is
identified with the so-called pseudogap phase in high T$_{c}$
cuprates. Superconductivity results from condensation of holons
$\langle b_{i1} \rangle \not= 0$ and $\langle b_{i2} \rangle = 0$
due to $i a_{i0}^{k} = \varphi \delta_{k3} \not= 0$ in the SF
phase while the Fermi liquid state appears from the SM phase in
the same way as the superconducting phase.

Low-energy physics and the stability of each phase should be
investigated beyond the mean-field description, quantum fluctuations
being introduced and an effective field theory being constructed.
Considering quantum fluctuations $U_{ij}^{SM} = - i \chi e^{i
a_{ij}^{k} \tau_{k} }$ in the SM phase, we can explain its low-energy
physics by an SU(2) gauge theory
\begin{eqnarray}
{\cal L}_{eff} &=&
\psi_{\alpha}^{\dagger} (\partial_{\tau} - ia_{\tau}^{k} \tau_{k}
) \psi_{\alpha} + \frac{1}{2m_{\psi}} |(\partial_{i} - i a_{i}^{k}
\tau_{k} )\psi_{\alpha}|^{2} \cr
&+& h^{\dagger}(\partial_{\tau}
- \mu - ia_{\tau}^{k}\tau_{k})h + \frac{1}{2m_{h}} |(\partial_{i}
- ia_{i}^{k}\tau_{k})h|^{2} \cr
& + & \frac{1}{4g^{2}}[\partial_{\mu} a_{\nu}^{k} -
\partial_{\nu} a_{\mu}^{k} - g \epsilon_{klm} a_{\mu}^{l}
a_{\nu}^{m}]^{2},
\end{eqnarray}
where the time and space components of the SU(2) gauge fields
arise from the Lagrange multiplier field and phase of the order
parameter matrix, respectively. $g$ stands for an effective
coupling constant. In this effective field theory the spinons
interact with the holons via SU(2) gauge fluctuations. Such gauge
interactions have been proposed as the source for the anomalous
transport in the SM phase~\cite{SM_U1GT}. However, it is not
enough to treat gauge fluctuations perturbatively in order to
simulate the Gutzwiller projection~\cite{Mudry_Confinement}.
Moreover, such an approach based on deconfinement cannot recover
the emergence of electron excitations at low temperatures without
the Anderson-Higgs mechanism \cite{Lee_Nagaosa_Wen}.

Defining the covariant derivative as
\begin{equation}
D_{\mu} = \partial_{\mu} -
i \phi \tau_{3} \delta_{\mu \tau} - ia_{\mu}^{k} \tau_{k}, \nonumber
\end{equation}
where $\phi$ is the mean-field part of the gauge
field associated with the Polyakov-loop parameter, and
incorporating quantum fluctuations $a_{\mu}^{k}$, we write down an
effective PNJL model for the matter sector
\begin{eqnarray}
\label{eq:3}
{\cal L}_{PNJL}^{M} &=& \psi_{\alpha}^{\dagger}
(\partial_{\tau} - i \phi \tau_{3}) \psi_{\alpha} +
\frac{1}{2m_{\psi}} | \partial_{i} \psi_{\alpha}|^{2} \cr
& + & h^{\dagger}(\partial_{\tau} - i \phi \tau_{3} - \mu )h +
\frac{1}{2m_{h}} | \partial_{i} h|^{2} \cr
& +& g_{\psi} \psi_{\alpha n}^{\dagger} \psi_{\alpha p} \psi_{\beta
  p}^{\dagger} \psi_{\beta n} + g_{c} \psi_{\alpha n}^{\dagger}
\psi_{\alpha p} h^{\dagger}_{p} h_{n},
\end{eqnarray}
where interactions between the spinons and holons are assumed to
be local. This local approximation is well utilized in the QCD
context, realizing SB$\chi$S successfully~\cite{PNJL_Models}. The
local current-current interactions are expected to be irrelevant
in the renormalization group sense, thus neglected for simplicity.

The spinon-exchange interaction, the first term of Eq.~(\ref{eq:3}) in
the last line, can be ignored in the SM phase while the electron
resonance term will be allowed as quantum corrections, later.
Then, we are led to the typical PNJL expression for the matter sector
%\begin{widetext}
\begin{eqnarray}
  \label{eq:4}
F_{M}^{SM}[\Phi,\mu;\delta,T] &=& -\frac{N_{s}}{\beta}
\sum_{k} \ln \left( 1 + 2 \Phi e^{- \beta \frac{k^{2}}{2m_{\psi}}
} + e^{- 2\beta \frac{k^{2}}{2m_{\psi}} } \right) \cr
&&\hspace{-2.5cm} +  \frac{1}{\beta} \sum_{q} \ln \left( 1 - 2 \Phi
e^{- \beta (\frac{q^{2}}{2m_{h}} - \mu )} + e^{- 2 \beta
(\frac{q^{2}}{2m_{h}} - \mu )}\right) + \mu \delta,
\end{eqnarray}
%\end{widetext}
where $\mu$ denotes the chemical potential and $\Phi = \cos \beta
\phi$ represents the Polyakov-loop parameter. Minimizing the free
energy with respect to $\Phi$, one always finds $\Phi = 1$, so
that Eq. (\ref{eq:4}) is reduced to the deconfined theory. Matter
fluctuations favor deconfinement as expected.

For the gauge sector, one can derive an effective theory of the
Polyakov-loop order parameter from pure Yang-Mills theory,
integrating over quantum fluctuations.
Unfortunately, the gauge free energy from one-loop approximation
always gives rise to $\Phi = 1$, i.e., deconfinement~\cite{Weiss}.
It is necessary to take quantum fluctuations into account in a
non-perturbative way. Such a procedure is not known yet, and we
construct an effective free energy as follows
\begin{eqnarray}
  \label{eq:5}
F_{G}[\Phi;T] = A_{4} T^{3} \Bigl\{ \frac{A_{2} T_{0}}{A_{4}}
\Bigl( 1 - \frac{T_{0}}{T} \Bigr) \Phi^{2} - \frac{A_{3}}{A_{4}}
\Phi^{3} + \Phi^{4} \Bigr\} ,
\end{eqnarray}
where the constants $A_{i=2,3,4}$ are positive definite, and $T_{0}$ is
identified with the critical temperature for the
confinement-deconfinement transition (CDT). Since the CDT is known
as the first order from the lattice simulation~\cite{PNJL_Models},
the cubic-power term with a negative constant is introduced such
that $\Phi = 0$ in $T < T_{0}$ while $\Phi = 1$ in $T > T_{0}$,
corresponding to the center symmetry ($Z_{2}$)
breaking~\cite{Fukushima}.

The resulting PNJL free energy is obtained as
\begin{equation}
\label{eq:6}
F_{PNJL}^{SM}[\Phi,\mu;\delta,T] \;=\; F_{M}^{SM}[\Phi,\mu;\delta,T] +
F_{G}[\Phi;T] .
\end{equation}
The CDT is driven by the gauge sector while
the matter fluctuations turn the first order transition into the
confinement-deconfinement crossover (CDC) because the $Z_{2}$ center
symmetry is explicitly broken in the presence of matters, so that the
Polyakov-loop does not become an order parameter in a rigorous
sense~\cite{Fukushima}. One may regard this PNJL construction as our
view point for the present problem according to experiments
\cite{ARPES}. An important point is that confinement changes both
spinon and holon spectra completely, allowing electron excitations,
although feedback effects of matters to the Polyakov-loop parameter
are not relevant.

\begin{figure}[ht]
%\vspace{0.5cm}
\centerline{\includegraphics[scale=0.77]{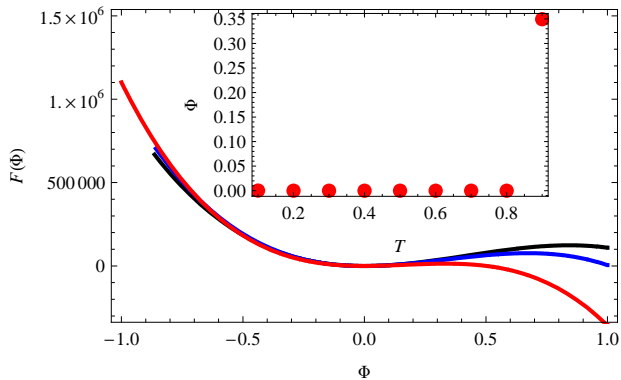} }
\caption{
(Color online) The effective PNJL free energy as a function of the
Polyakov-loop parameter with $T < T_{CD}$ (Black), $T = T_{CD}$
(Blue), and $T > T_{CD}$ (Red).
Inset: The Polyakov-loop parameter as a function of temperature
scaled with $T_{0}$.
}
\label{fig:1} %\vspace{-0.5cm}
\end{figure}

Figure~\ref{fig:1} shows the free energy as a function of the
Polyakov-loop parameter for various temperatures, where $\Phi = 0$
in $T < T_{CD}$ (Black) and $\Phi = 1$ in $T > T_{CD}$ (Red) as it
should be. The blue curve is drawn at $T = T_{CD}$. Here, $T_{CD}$ is
the CDC temperature in the presence of matters, smaller than
$T_{0}$ because matters favor the deconfinement. An interesting
point is that the chemical potential of a negative value is much
larger in the confinement phase than in the deconfinement phase,
consistent with confinement. The inset displays the Polyakov-loop
parameter that starts to appear around $T < T_{0}$.

An interesting result in the mean-field approach of the PNJL model
is that condensation of bosons is not allowed, since the expression
for the boson sector cannot reach the zero value
because of $0 \leq \Phi < 1$ except for $\Phi = 1$. In
other words, Higgs phenomena are not compatible with confinement
in this description, consistent with the previous field-theoretic
result~\cite{Higgs_Confinement}.

Since the Higgs phase is not allowed in the presence of the
Polyakov-loop parameter, an immediate issue is how to describe the
Fermi liquid phase, usually given by condensation of holons. We
will examine the electron self-energy in the confinement phase,
certainly recovering the Fermi-liquid self-energy proportional to
$\omega^{2}$ with frequency $\omega$ below a certain temperature
associated with the holon chemical potential.

We repeat a similar study for the SF phase in which the spinon
sector is described by the relativistic spectrum. The evolution of the
Polyakov-loop parameter tends to be almost the same as the case of the
SM phase. Note that the condensation of holons is also not
allowed. Thus, the present framework presents a possible new mechanism
of superconductivity in the presence of confinement instead of the
holon condensation in the deconfinement phase.
%%%%%%%%%%%%%%%%%%%%%%%%%%%%%%%%%%%%%%%%%%%%%%%%%%%%%%%%%%%%%%%%%%%%%%%%
%%%%%%%%%%%%%%%%%%%%%%%%%%%%%%%%%%%%%%%%%%%%%%%%%%%%%%%%%%%%%%%%%%%%%%%%

The central question of the present work is on the fate of the
spinon and holon when the Polyakov-loop parameter vanishes. The
spinon-holon coupling term in Eq.~(\ref{eq:3}) can be expressed as
follows
\begin{equation}
\mathcal{S}_{el} \;=\; \int_{0}^{\beta} d \tau \int
d^{2} r \left( \psi_{\sigma n}^{\dagger} h_{n} c_{\sigma} +
c_{\sigma}^{\dagger} h^{\dagger}_{p} \psi_{\sigma p} -
\frac{1}{g_{c}} c_{\sigma}^{\dagger} c_{\sigma} \right)  ,\nonumber
\end{equation}
where $\sigma$ and $n(p)$ represent spin and SU(2) indices,
respectively. Since the Grassmann variable $c_{\sigma}$ carries
exactly the same quantum numbers with the electron, one may
identify it as the Hubbard-Stratonovich field $c_{\sigma}$.
The effective coupling constant $g_{c}$ plays a role of the chemical
potential for electrons. Note that the Fermi surface of the electrons
differs from that of the spinons in principle.

One can introduce the quantum corrections self-consistently in the
Luttinger-Ward functional approach \cite{LW} in which only
planar diagrams are taken into account, ignoring vertex
corrections~\cite{Kim_LW}. We arrive at the self-consistent equations
for self-energies
\begin{eqnarray}
\label{eq:8}
&&\Sigma^{c}_{\sigma\sigma}(k,i\omega) = -
\frac{1}{\beta} \sum_{i\Omega} \sum_{q} G^{h}_{p'p}(q,i\Omega)
G^{\psi}_{\sigma\sigma,pp'} (k-q,i\omega-i\Omega) , \cr
&&\Sigma^{\psi}_{\sigma\sigma,pp'} (k,i\omega) = - \frac{1}{\beta}
\sum_{i\Omega} \sum_{q} G^{c}_{\sigma\sigma} (k+q,i\omega+i\Omega)
G^{h}_{p'p}(q,i\Omega) , \cr
&&\Sigma^{h}_{pp'}(q,i\Omega) =
\frac{1}{\beta} \sum_{i\omega} \sum_{k} G^{c}_{\sigma\sigma}
(k+q,i\omega+i\Omega) G^{\psi}_{\sigma\sigma,pp'} (k,i\omega) ,
\end{eqnarray}
where the Green's functions for the electron, the spinon, and the
holon are given as
\begin{eqnarray}
- G^{c -1}_{\sigma\sigma} (k,i\omega) &=&
\Sigma^{c}_{\sigma\sigma}(k,i\omega) - g_{c}^{-1} , \cr
-G^{\psi -1}_{\sigma\sigma,pp'} (k,i\omega) &=& - i (\omega + p \phi)
\delta_{pp'} + \frac{k^{2}}{2m_{\psi}} \delta_{pp'}
+ \Sigma^{\psi}_{\sigma\sigma,pp'} (k,i\omega) , \cr
- G^{h
-1}_{pp'}(q,i\Omega) &=& [- i ( \Omega + p \phi) - \mu] \delta_{pp'}
+ \frac{q^{2}}{2m_{h}} \delta_{pp'} \cr
&& + \Sigma^{h}_{pp'}(q,i\Omega) ,
\end{eqnarray}
respectively. These equations were intensively discussed in the
context of heavy fermions \cite{Kim_LW,Pepin_Paul} without confinement
due to the Polyakov-loop parameter.

It is natural that the spectral function of the spinon should not
be reduced to the delta function even if the self-energy
correction is ignored owing to the presence of the background
potential $\phi$. That of the holon also features a broad
structure even at the zero frequency because of the Polyakov-loop
parameter. It indicates that both the spinon and holon are not
well-defined excitations in the confinement phase. On the other
hand, the electron as a spinon-holon composite exhibits a rather
sharp peak, since the imaginary part of their self-energy vanishes
in the zero frequency limit in spite of no pole structure in the
Green's function.

The holon self-energy is found to be of the standard form in two
dimensions
%\begin{widetext}
  \begin{eqnarray}
&&\Sigma_{p}^{b}(q,i\Omega) - \Sigma_{p}^{b}(q,0) \cr
&&= - \frac{\rho_{c} }{i(\alpha-1)}
\Bigl\{\tan^{-1}\Bigl(\frac{i\Omega+ip\phi- v_{F}^{c} q^{*} +
v_{F}^{c} q}{- i\Omega - i p\phi + v_{F}^{c} q^{*} + v_{F}^{c}
q}\Bigr) \cr
&&- \tan^{-1}\Bigl(\frac{i\Omega+ip\phi-\alpha
v_{F}^{c} q^{*} + \alpha v_{F}^{c} q}{- i\Omega - i p\phi + \alpha
v_{F}^{c} q^{*} + \alpha v_{F}^{c} q}\Bigr) \Bigr\}
\label{eq:9}
 \end{eqnarray}
%\end{widetext}
except for $i\Omega \rightarrow i\Omega+ip\phi$. $\rho_{c}$ is the 
density of states for the confined electron, and $v_{F}^{c}$
stands for the corresponding Fermi velocity. $\alpha$ denotes the 
ratio of the electron band mass to the spinon one, given as almost
unity. $q^{*}$ designates the Fermi-momentum mismatch between the
confined electron and spinon.

%%%%%%%%%%%%%%%%%%%%%%%%%%%%%%%%%%%%%%%%%%%%%%%%%%%%%%%%%%%%%%%%%%%%%%%%
\begin{figure}[htp]
\centerline{\includegraphics[scale=0.58]{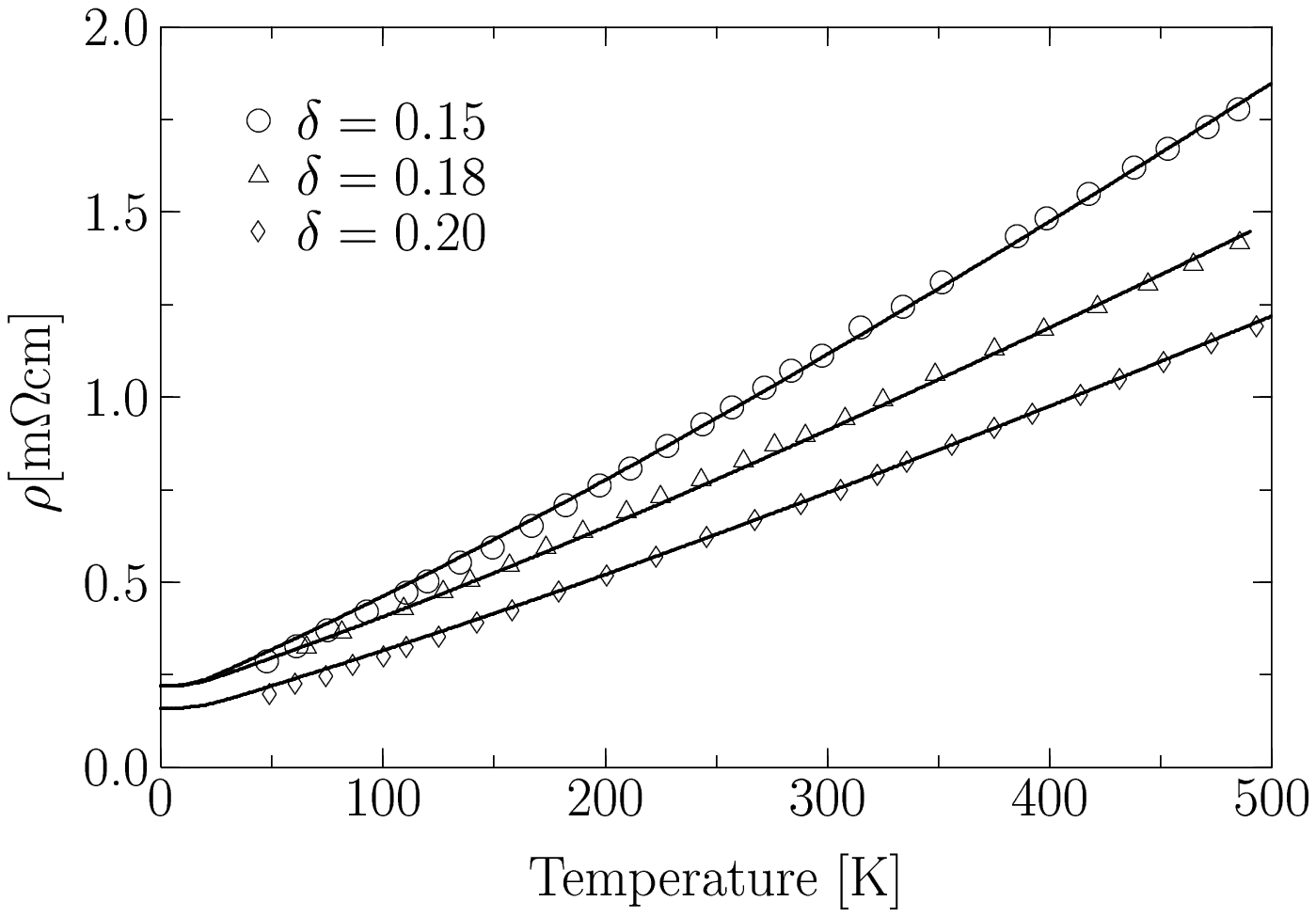}}
\caption{ (Color online) The electrical resistivity
(Ref. \cite{Data}) with parameter $\mathcal{C}$ fitted.}
\label{fig2}
\end{figure}
%%%%%%%%%%%%%%%%%%%%%%%%%%%%%%%%%%%%%%%%%%%%%%%%%%%%%%%%%%%%%%%%%%%%%%%%
Inserting Eq.~(\ref{eq:9}) into the electron self-energy equation,
we can find its explicit form. An important energy scale is given
by the holon chemical potential $\mu$. In $T > |\mu|$ holon
dynamics is described by the dynamical exponent $z = 3$, resulting
from the Landau damping of the electron and
spinon~\cite{Kim_LW,Pepin_Paul}. The imaginary part of the
self-energy turns out to be proportional to $T^{2/3}$, since the
confined electrons are scattered with such $z = 3$ dissipative
modes~\cite{Kim_LW,Pepin_Paul}. On the other hand, the holon
excitations have gaps in $T < |\mu|$, recovering the Fermi liquid.
Thus, the Fermi liquid appears as the confinement phase rather
than the Higgs in the PNJL approach.

We fit the resistivity data~\cite{Data} for optimally doped
cuprates. The relaxation time differs from the transport time, and the
back scattering contribution is factored out by vertex corrections,
corresponding to $T^{2/3}$ for two dimensional $z = 3$
fluctuations \cite{Kim_TR}. Then, the final expression can be written
as
\begin{equation}
  \label{eq:10}
\rho_{el}(T) \;=\; \rho_{0} + \mathcal{C} \Bigl( N_{s} \rho_{c}
\frac{v_{F}^{c2}}{3} \Bigr)^{-1} T^{2/3} \Im \Sigma_{c}(T) ,
\end{equation}
where $\rho_{0}$, $\mathcal{C}$, and $N_{s}$ denote, respectively,
the residual resistivity due to disorder, the strength for vertex
corrections, and the spin degeneracy. $\rho_0$ and $\mathcal{C}$
are free parameters. Interestingly, the residual resistivity turns
out to be almost constant for several hole doped samples near the
optimal doping. Thus, we have practically only one free parameter,
i.e. $\mathcal{C}$. As shown in Fig. 2, the results are in
remarkable agreement with the data, which supports our confinement
scenario.

In the present work, we have introduced an effect of confinement into
the gauge theory approach for strongly correlated electrons, taking
into account the Ployakov-loop parameter. We were able to identify the
coherence temperature at which confinement of the spinon and holon,
yielding electron excitations, emerges. Remarkably, such
electron excitations are not fully coherent, which is consistent with
the non-Fermi liquid physics observed in the optimally doped region.
It was demonstrated explicitly by fitting the resistivity data
(Fig. 2).  A unique feature is that the Higgs mechanism does not arise
in the presence of the Polyakov-loop parameter, which implies a
possible novel mechanism  of superconductivity beyond the existing
gauge theoretical framework.

It will be of great interest to apply the PNJL scheme to the
spin-liquid theory \cite{Spin_Liquid} and Kondo breakdown scenario
\cite{Pepin_Paul} for heavy fermions. In the former the
crossover behavior from spin $1$ excitations to spin $1/2$ can be
investigated while this new mechanism for heavy fermions may occur in
the latter.

K.-S. Kim was supported by the National Research Foundation of
Korea (NRF) grant funded by the Korea government (MEST) (No.
2009-0074542). The present work (H.-Ch. Kim) is also supported by
Basic Science Research Program through the NRF funded by the MEST
(No. 2009-0073101).

\end{document}